\documentclass[aps,prc,twocolumn,showpacs,floatfix,nofootinbib,preprintnumbers,superscriptaddress,amsmath,amssymb]{revtex4-2}
\usepackage[T1]{fontenc}
\usepackage[utf8]{inputenc}
\usepackage{graphicx}
\usepackage{epsfig}
\usepackage{bm}
\usepackage{color}
\usepackage{float}
\usepackage{dcolumn}
\usepackage{txfonts}

\begin{document}

%--------
\title{Decay study of the most neutron-deficient Zn isotopes with the Warsaw Optical TPC detector}
%--------
\date{\today}

\author{A.~Kubiela}
\email{adam.kubiela@fuw.edu.pl}
\affiliation{Faculty of Physics, University of Warsaw, 02-093 Warszawa, Poland}
%
%\author{O.B.~Tarasov}
%\affiliation{National Superconducting Cyclotron Laboratory, Michigan State University, East Lansing, Michigan 48824, USA}
%
\author{D.-S.~Ahn}
\affiliation{RIKEN Nishina Center, 2-1 Hirosawa, Wako, Saitama 351-0198, Japan}
\author{H.~Baba}
\affiliation{RIKEN Nishina Center, 2-1 Hirosawa, Wako, Saitama 351-0198, Japan}
\author{A.~Bezbakh}
\affiliation{Flerov Laboratory of Nuclear Reactions, JINR, 141980 Dubna, Russia}
\author{A.A.~Ciemny}
\affiliation{Faculty of Physics, University of Warsaw, 02-093 Warszawa, Poland}
\author{W.~Dominik}
\affiliation{Faculty of Physics, University of Warsaw, 02-093 Warszawa, Poland}
\author{N.~Fukuda}
\affiliation{RIKEN Nishina Center, 2-1 Hirosawa, Wako, Saitama 351-0198, Japan}
\author{A.~Giska}
\affiliation{Faculty of Physics, University of Warsaw, 02-093 Warszawa, Poland}
\author{R.~Grzywacz}
\affiliation{Department of Physics and Astronomy, University of Tennessee, Knoxville, Tennessee 37996, USA}
\author{V.~Guadilla}
\affiliation{Faculty of Physics, University of Warsaw, 02-093 Warszawa, Poland}
\author{Y.~Ichikawa}
\affiliation{RIKEN Nishina Center, 2-1 Hirosawa, Wako, Saitama 351-0198, Japan}
\affiliation{Department of Physics, Kyushu University, 744 Moto-oka, Nishi, Fukuoka, Fukuoka 819-0395, Japan}
\author{Z.~Janas}
\affiliation{Faculty of Physics, University of Warsaw, 02-093 Warszawa, Poland}
\author{Ł.~Janiak}
\affiliation{National Centre for Nuclear Research, 05-400 Otwock, Świerk, Poland }
\author{G.~Kamiński}
\affiliation{Heavy Ion Laboratory, University of Warsaw, 02-093 Warsaw, Poland}
\author{K.~Kawata}
\affiliation{RIKEN Nishina Center, 2-1 Hirosawa, Wako, Saitama 351-0198, Japan}
\affiliation{Center for Nuclear Study, University of Tokyo, 7-3-1 Hongo, Bunkyo, Tokyo 113-0033, Japan}
\author{T.~Kubo}
\affiliation{RIKEN Nishina Center, 2-1 Hirosawa, Wako, Saitama 351-0198, Japan}
\author{M.~Madurga}
\affiliation{Department of Physics and Astronomy, University of Tennessee, Knoxville, Tennessee 37996, USA}
\author{C.~Mazzocchi}
\affiliation{Faculty of Physics, University of Warsaw, 02-093 Warszawa, Poland}
\author{H.~Nishibata}
\affiliation{RIKEN Nishina Center, 2-1 Hirosawa, Wako, Saitama 351-0198, Japan}
\affiliation{Department of Physics, Kyushu University, 744 Moto-oka, Nishi, Fukuoka, Fukuoka 819-0395, Japan}
\author{M.~Pomorski}
\affiliation{Faculty of Physics, University of Warsaw, 02-093 Warszawa, Poland}
\author{Y.~Shimizu}
\affiliation{RIKEN Nishina Center, 2-1 Hirosawa, Wako, Saitama 351-0198, Japan}
\author{N.~Sokołowska}
\affiliation{Faculty of Physics, University of Warsaw, 02-093 Warszawa, Poland}
\author{D.~Suzuki}
\affiliation{RIKEN Nishina Center, 2-1 Hirosawa, Wako, Saitama 351-0198, Japan}
\author{H.~Suzuki}
\affiliation{RIKEN Nishina Center, 2-1 Hirosawa, Wako, Saitama 351-0198, Japan}
\author{P.~Szymkiewicz}
\affiliation{AGH University of Science and Technology, Faculty of Physics and Applied Computer Science, 30-059 Krakow, Poland}
\author{A.~Świercz}
\affiliation{AGH University of Science and Technology, Faculty of Physics and Applied Computer Science, 30-059 Krakow, Poland}
\author{M.~Tajima}
\affiliation{RIKEN Nishina Center, 2-1 Hirosawa, Wako, Saitama 351-0198, Japan}
\author{A.~Takamine}
\affiliation{RIKEN Nishina Center, 2-1 Hirosawa, Wako, Saitama 351-0198, Japan}
\author{H.~Takeda}
\affiliation{RIKEN Nishina Center, 2-1 Hirosawa, Wako, Saitama 351-0198, Japan}
\author{Y.~Takeuchi}
\affiliation{RIKEN Nishina Center, 2-1 Hirosawa, Wako, Saitama 351-0198, Japan}
\affiliation{Department of Advanced Sciences, Hosei University}
\author{C.R.~Thornsberry}
\affiliation{Department of Physics and Astronomy, University of Tennessee, Knoxville, Tennessee 37996, USA}
\author{H.~Ueno}
\affiliation{RIKEN Nishina Center, 2-1 Hirosawa, Wako, Saitama 351-0198, Japan}
\author{H.~Yamazaki}
\thanks{deceased on 15 July 2024}
\affiliation{RIKEN Nishina Center, 2-1 Hirosawa, Wako, Saitama 351-0198, Japan}
\author{R.~Yokoyama}
\affiliation{Department of Physics and Astronomy, University of Tennessee, Knoxville, Tennessee 37996, USA}
\author{K.~Yoshida}
\affiliation{RIKEN Nishina Center, 2-1 Hirosawa, Wako, Saitama 351-0198, Japan}
\author{M.~Pf\"utzner}
\email{pfutzner@fuw.edu.pl}
\affiliation{Faculty of Physics, University of Warsaw, 02-093 Warszawa, Poland}

\begin{abstract}
Results of decay studies of nuclei in the vicinity of $^{54}$Zn, which is the most
neutron-deficient isotope of zinc and undergoes ground-state two-proton
radioactivity (\emph{2p}), are presented. The measurements were performed 
with a gaseous time projection chamber with optical readout which allowed us
to record tracks of protons emitted in the decays. A new method of data analysis 
was used to reconstruct energies and emission angles of low-energy protons
that were stopped within the active volume of the chamber. Half-lives and
branching ratios for $\beta$-delayed proton emission channels were determined
for $^{56}$Zn, $^{55}$Zn, and $^{55}$Cu. The $\beta$-delayed emission of
two protons for $^{55}$Zn was observed for the first time. Five events of
\emph{2p} radioactivity of $^{54}$Zn were detected and reconstructed. 
The distribution of the opening angle between momenta of the two protons 
is consistent with the findings published in [Ascher et al. PRL 107, 102502 (2011)].
The combination of all results suggests a flat angular distribution, in 
contrast to the one measured for $^{45}$Fe. 

\end{abstract}

\maketitle

\section{Introduction}

The study of nuclei at the proton dripline and beyond is one of the current frontiers
of nuclear physics. Although this edge of the nuclidic chart is better explored than
the neutron-rich side, many questions concerning proton-dripline nuclei remain to be
answered. The nuclear properties in this region result from the interplay between
large $\beta$-decay $Q$ values, low or negative proton separation energies, and the
confining effects of the Coulomb barrier. The emerging characteristic phenomena
include a variety of $\beta$-delayed particle emission channels, proton radioactivity,
and two-proton radioactivity \cite{Blank:2008,Pfutzner:2012,Pfutzner:2023}.
These decay modes provide essential input to the nuclear structure modeling in this region.
In addition, the properties of these nuclei are of key importance
for the description of the astrophysical $rp$-process~\cite{Schatz:1998}.

Two-proton radioactivity occurs for unbound even-Z nuclei for which the single-proton
separation energy is positive or very small \cite{Pfutzner:2012}. The complete investigation
of a two-proton emitting nucleus involves, in addition to the measurements of the
half-life, the decay energies and the branching ratios, also the study of the correlations
of momenta of the two emitted protons. The latter observable offers a new and unique
insight into the structure of exotic, proton-unbound nuclei.
This information, however, is still scarce and limited to a few cases \cite{Pfutzner:2023}.
The main difficulty in obtaining such data is the production of 2p-decaying nuclei in amounts
sufficient for statistically significant correlations studies.

Of particular interest is the region of proton unbound nuclei around the
proton number $Z=28$.
Three cases of 2p radioactivity were identified here: $^{45}$Fe
($Z=26$)~\cite{Pfutzner:2002,Giovinazzo:2002}, $^{48}$Ni ($Z=28$)~\cite{Pomorski:2011},
and $^{54}$Zn ($Z=30$)~\cite{Blank:2005}.
An exciting possibility arises, that the detailed study of two-proton correlations
for these three cases may reveal whether the number $Z=28$ still
corresponds to a closed proton shell so far from the stability line \cite{Miernik:2009}. 

Motivated by the excellent performance of the BigRIPS separator \cite{Kubo:2003,Kubo:2012}, 
and in particular by the large intensity of the $^{78}$Kr beam available at the 
Radioactive Isotope Beam Factory (RIBF) of the RIKEN Nishina Center,
we produced the 2p-decaying nucleus $^{54}$Zn in $^{78}$Kr
fragmentation. First results of this study, on the production cross section for
the three most exotic zinc isotopes, $^{54-56}$Zn, together with a compilation of
the measured cross sections for the most neutron-deficient zinc, germanium, selenium,
and krypton isotopes, and with a comparison to model predictions, were published
in Ref.~\cite{Kubiela:2021}. Here, we report on the results of spectroscopic
decay studies of $^{54}$Zn and neighboring nuclei. We focus on decays with emission
of protons by employing the Warsaw gaseous TPC detector with optical readout (OTPC).

%======================================================================================================

\section{Experimental techniques}
\subsection{Production and identification of ions}

The experiment was carried out at the RIBF facility. 
The nuclei of interest were produced using a $^{78}$Kr primary
beam at 345~MeV/nucleon impinging on a 10 mm-thick beryllium target and separated
with the help of the large-acceptance two-stage fragment separator BigRIPS \cite{Kubo:2003,Kubo:2012}.
Two aluminum wedge-shaped degraders were used: 4~mm-thick and 1.5~mm-thick
at the momentum dispersive focal planes F1 and F5, respectively.
Ions coming to the final focal plane of the separator (F7) were identified
in flight by a set of standard BigRIPS detectors and the $\Delta E$-TOF-$B \rho$
method \cite{Fukuda:2013}. The time of flight (TOF) was measured by two 500 $\mu$m
thick plastic scintillators mounted at the achromatic foci F3 and F7. The magnetic
rigidity ($B \rho)$ was determined from a set of particle tracking detectors --- position-sensitive parallel-plate avalanche counters (PPAC) --- located
at F3, F5, and F7 focal planes. The energy loss ($\Delta E$) was measured
using the ionization chamber placed at the F7 focus. Ions were
further transferred, via the ZeroDegree spectrometer to the final plane F11
where the OTPC detection system was located. In front of the OTPC detector,
a 300 $\mu$m-thick Si detector was mounted which provided additional
energy-loss information.

Data from TOF and $\Delta E$ detectors, except for the silicon detector in front of the OTPC, were recorded by BigRIPS data acquisition system. The TOF signals from the detectors located at F3 and F7, and signals from the silicon detector in front of the OTPC were also recorded by the OTPC acquisition system. Both systems ran independently, but were synchronized by the same clock, so events registered in one system could be identified in the other.
The identification of ions was achieved following the methods presented in Ref.~\cite{Fukuda:2013} and was described in more detail in Ref.~\cite{Kubiela:2021}. The second identification of particles arriving to the OTPC, was performed using TOF signals from the detectors located at F3 and F7 and a signal from the silicon detector mounted at F11, just in front of the OTPC window.
These identification signals were used to construct a trigger signal activating the OTPC acquisition system, so only data for incoming ions of interest in a given setting were stored on disk.

Data were collected using three main settings of the separator, optimized for the
transmission of $^{56}$Zn, $^{55}$Zn, and $^{54}$Zn, respectively. For the cases of $^{56}$Zn and
$^{54}$Zn, two variants of the setting with slightly different opening of the F2 slit were used.
The relevant parameters of these settings were given in Table I of Ref.~\cite{Kubiela:2021}.
The average beam current was varying from about 8 pnA for $^{56}$Zn to
about 240 pnA for $^{54}$Zn.

\subsection{The OTPC detector}

The optical time projection chamber (OTPC) was developed at the University of Warsaw
specifically to study very rare decay modes with emission of charged particles,
such as 2p radioactivity \cite{Pomorski:2014,Ciemny:2022}.
The OTPC is a gas-filled detector with an active volume 33 cm deep, 20 cm wide, and 21 cm high. In this study a gas mixture of 69\% argon, 29\% helium and 2\% tetrafluoromethane (CF$_4$) at atmospheric pressure was used. 
The selection of this mixture was the result of a compromise between efficient stopping of implanted ions and  
good visibility of tracks made by protons having energy of about 1~MeV. 
The active volume is immersed in an homogenous vertical electric field (143 V/cm). The ions pass first through a variable aluminum degrader and enter the chamber horizontally, perpendicular to the field lines, through a thin kapton window.
The degrader is set to maximize the number of ions of interest that stop in the active volume and subsequently decay. The ionization electrons generated by the interaction of a heavy ion and its charged-particle decay products drift at a constant velocity in the electric field towards an amplification structure based on a set of four gas-electron multiplier (GEM) foils \cite{Sauli:1997} and the anode.
A special grid electrode, mounted before the GEM section, is used to control the charge 
to be amplified. In this way, a low sensitivity is maintained when highly charged ions enter
the chamber. While waiting for the decay of the stopped ion, the electrode potential is switched
to the high sensitivity mode. 
In this experiment, light generated at the final stage was recorded by two CCD cameras and a photo-multiplier tube (PMT).
One camera (the implantation camera) was set to photograph the track of a triggering ion, while the exposition of the second one (the decay camera) was started after the trigger to record the decay event. The resolution of both cameras was $512 \times 512$ pixels, which translates to 0.66 mm/pixel horizontal resolution on the image.
The exposure times of the decay camera were 120 ms in the $^{56}$Zn and $^{55}$Zn settings, and 20 ms in the $^{54}$Zn setting. The implantation camera was sensitive up to 1 ms time before the trigger.
The PMT connected to the oscilloscope was recording the light output during the event. The sampling rate of the PMT oscilloscope was 100 MSa/s. 
To monitor the synchronization between the stored CCD image and the PMT waveform a system of
light diodes was used. A row of 16 diodes was mounted close to the bottom edge of the active volume
in such a way that their light was visible by both the CCD camera and the PMT. Their light and flash time pattern (on/off) was used to code a binary number, advanced by 1 at each OTPC trigger.

The electron drift velocity in the electric field of the OTPC was monitored during the entire experiment with a separate drift velocity detector connected to the OTPC gas line. The measured average value of
the drift velocity was 19.36(28) mm/$\mu$s. The sampling rate of the PMT oscilloscope, combined with the drift velocity, gives the vertical spatial resolution of 0.19 mm/sample. Temperature and pressure influencing the gas density were monitored with sensors in proximity to the OTPC.

%\begin{figure}[ht]%fig1
%\begin{center}
%\includegraphics[scale=0.55]{Fig_new.pdf}
%\end{center}
%\vspace*{-4mm}
%\caption{(Color online) Particle identification plots showing the mass-to-charge ratio $A/Q$ and the atomic %number $Z$ for the three zinc isotopes investigated. The plots for $^{56}$Zn (a) and for $^{54}$Zn (c) %correspond
%to settings detailed in the second and the fourth row of Table I, respectively.
%}
%\label{fig1}
%\end{figure}

\section{Analysis of data}

\subsection{Event selection, branching ratio and half life}
% The procedure for establishing branching ratios and half lives of produced ions was as follows.
In the offline data analysis, for each setting of BigRIPS, all saved events were inspected one by one.
First, the events were grouped according to the identity of the triggering ion.
Next, data from the implantation camera were used to check if the triggering ion did stop inside the active volume of the OTPC chamber.
Only events where the ion was stopped not closer than 1.5 cm from the detector walls were retained
for the further analysis. This condition was adopted to make sure that particles emitted in the
decay would be observed with 100\% efficiency.

Then, using combined data from the decay camera and the PMT, the decay events were counted and decay mode was identified. From the number of stopped ions, $N_0$, and the number of observed decay
events, $N_d$, the branching ratio for the given decay mode was determined:
\begin{equation}\label{}
  \begin{split}
    BR(d) = & \frac{N_d}{N_0 \, f} ,                                                     \\
    f =     & \,\lambda \int_{t_{\rm min}}^{t_{\rm max}} \exp(- \lambda \, t) \, dt \, ,
  \end{split}
\end{equation}
where $f$ is the correction factor for the finite observation time and $\lambda$ is the
decay constant of the ion of interest. The integration limits correspond to the settings of
the decay camera and amounted to $t_{\rm min} = 0.4$~ms, $t_{\rm max} = 120$~ms 
(in case of the $^{54}$Zn setting, $t_{\rm max} = 20$~ms). 

The decay time read from the PMT waveform was used to determine the mean decay lifetime using the maximum likelihood method. Again, taking into account the finite observation time, the maximum
likelihood function for $n$ decay events occurring at times $t_1, ... t_n$ reads:
\begin{equation}\label{}
  L(\tau; t_1,...t_n) = \frac{\lambda \, \prod_{i=1}^{n} \exp(- \lambda \, t_i)}{\exp(- \lambda \, t_{\rm min}) -  \exp(- \lambda \, t_{\max} )} .
\end{equation}
The maximum of numerically calculated $\ln L$ as a function of $\tau$ yielded the mean
lifetime and its uncertainty.

\subsection{Track reconstruction}
In case of $^{55}$Zn and $^{54}$Zn, emission of protons which were stopped inside
the active volume of the OTPC chamber was observed. The decay camera images and the PMT waveforms
corresponding to these decay events were used to fully reconstruct the tracks of emitted protons.

The track reconstruction, that is obtaining the energy of every particle emitted in a decay and its direction in space, was done by comparing the signals recorded by the CCD and the PMT with a theoretical model of an OTPC event.
In case of single-track events (one particle emitted with negligible recoil) it is often sufficient to determine the energy and the direction from the lengths of the two projections of the track, as observed on the CDD image and in the PMT waveform. The reconstruction of events with two or more tracks, however, requires taking into account also the distribution of the energy deposited along each track.
In this work, we adopted the reconstruction procedure based on the latter approach.

The measured light signals on the image and in the waveform are proportional to the electronic stopping power $dE/dx$ of a charged particle. Using the energy-loss data from the SRIM2013 code \cite{SRIM:2010} and knowing the OTPC gas mixture composition, together with the temperature and the pressure of the gas, the energy-deposition profile along the track for the particle with a given initial energy can be calculated. Then, the projections of this profile on both the horizontal plane (CCD image) and the vertical axis (PMT waveform) is determined based on the assumed polar and the azimuthal angles of the particle track in 3D space.
%The length of the horizontal projection is calculated from mm to pixels with a known pixel-to-mm ratio, and the length of a vertical track projection is calculated from mm to microseconds with an electron drift velocity.
To account for the diffusion of the ionization electrons, both projections of the energy-deposition profile are folded with Gaussian functions. Finally, overall scaling factors for signal intensities are applied.

In the analysis of $\beta$-delayed emission of a single proton, our model has 10 parameters:
the track lengths on the CCD image $L^{CCD}_1$ (horizontal projection) and in the PMT waveform $L^{PMT}_1$ (vertical projection), the azimuthal angle $\phi_1$ in the image plane, 
the coordinates of the decay position $x, y$ (CCD) and $z$ (PMT),
two widths of the Gaussian filters, and the scaling factors for both projections. 
Correspondingly, in case of a two-proton decay event, the model has 13 parameters with three additional ones, $L^{CCD}_2$, $L^{PMT}_2$ and $\phi_2$, for the second track.   
The values of the best fitting parameters to a given decay event were determined by finding the global minimum
of the following function $S$:
\begin{equation}
  \begin{split}
    S_{\rm CCD} &= \frac{1}{\hat{\sigma}_{\rm CCD}^2}\sum_{i,j} \left[CCD_{\rm exp}(i,j) - CCD_{\rm mod}(i,j)\right]^2 \, , \\
    S_{\rm PMT} &= \frac{1}{\hat{\sigma}_{\rm PMT}^2}\sum_{k} \left[PMT_{\rm exp}(k) - PMT_{\rm mod}(k)\right]^2 \, ,       \\
    S &= S_{\rm CCD} + S_{\rm PMT} \, ,
  \end{split}
\end{equation}
where $CCD(i,j)$ is the corresponding pixel value on the CCD image, $PMT(k)$ is the k-th value of the PMT waveform, the subscripts `exp' and `mod' refer to the data and the model, respectively, and $\hat{\sigma}_{\rm CCD}$ and $\hat{\sigma}_{\rm PMT}$ are the uncertainties of signal intensities due to noise.
Both of them were estimated by smoothing the corresponding
signal and comparing it with the original one \cite{Sokolowska:2024}:
\begin{equation}
  \hat{\sigma}_{\rm d}^2 = \frac{1}{N} \sum_{n=1}^{N} \left[ d_{\rm exp}(n) - d_{\rm smooth}(n) \right]^2 \, ,
\end{equation}
where $d_{\rm smooth}$ is the result of smoothing the $d_{\rm exp}$ data set with a Gaussian filter and
the summation goes for all points of the given data channel.
The width of the filter was chosen by testing the method on model events with a simulated noise.

The global minimum of $S$ was found by means of a direct search method, evaluating both $S_{\rm CCD}$ and $S_{\rm PMT}$ for every parameter combination in a reasonably broad parameter space.
%Scaling factors were not varied, they were calculated for every point in the parameter space by linear least squares %method.
Initial parameter values were chosen by manually inspecting the track ends on the image and the waveform, and choosing a typical track width for Gaussian filters.
Two searches were performed per event, first with a coarse, and later with a fine step sizes.
For the coarse-grain scan, the step sizes were set to 4 pixels/samples for the track lengths and the
decay position, 1.3 pixels/samples for the track widths, and 8 degrees for the azimuthal angle.
During the fine-grain scan, the step sizes were reduced by a factor of four.
The volume of the parameter space in the coarse-grain scan was taken large enough to eliminate the
possibility of missing the global minimum, while the space for the fine-grain scan was adjusted
based on the coarse search results.
The adopted direct search method is much slower than typical gradient-descend type methods for solving non-linear least-squares problems. However, it practically guarantees finding the global minimum,
regardless of the starting point, so it is very useful when analysing low-statistics data.
%\textcolor{blue}{Describe how we get energy from $L^{CCD}$ and $L^{PMT}$?}
\begin{figure*}[!ht]
  \centering
  \includegraphics[width=0.8\textwidth]{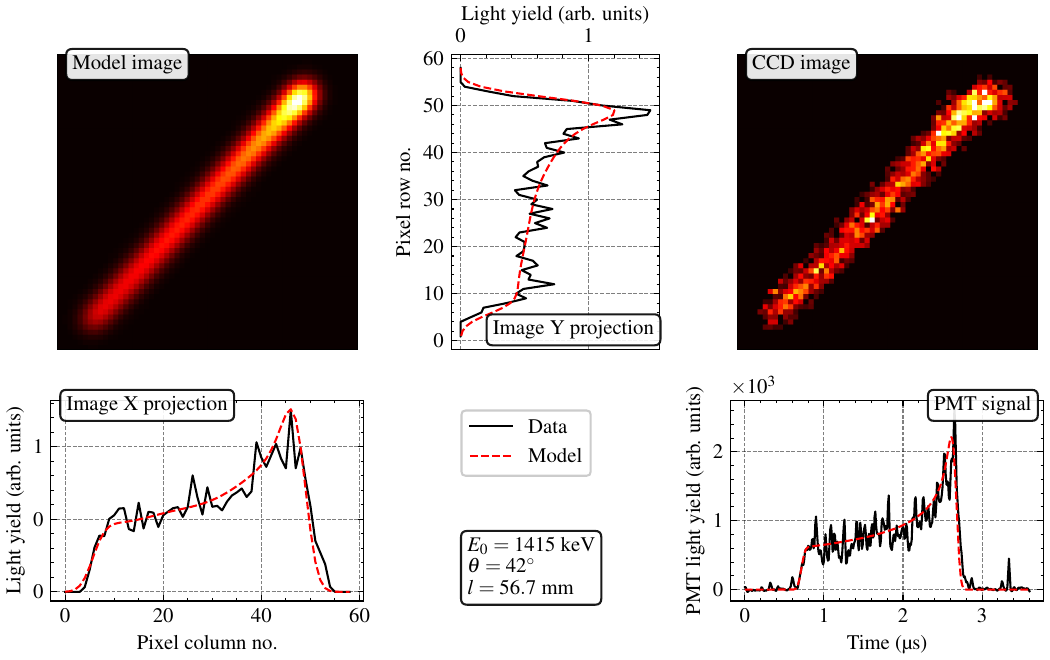}
  \caption{An example of the reconstruction of a $\beta$-delayed proton emission following the decay of $^{55}$Zn. $E_0$ is the initial kinetic energy of the proton, $\theta$ is the angle between the
  proton track and the horizontal plane, and $l$ is the length of the track in 3D.
  The scale origin for the time and for the pixel numbers was chosen arbitrarily.  }
  \label{fig:bpReconstruction}
\end{figure*}

\begin{figure*}[!ht]
  \centering
  \includegraphics[width=0.8\textwidth]{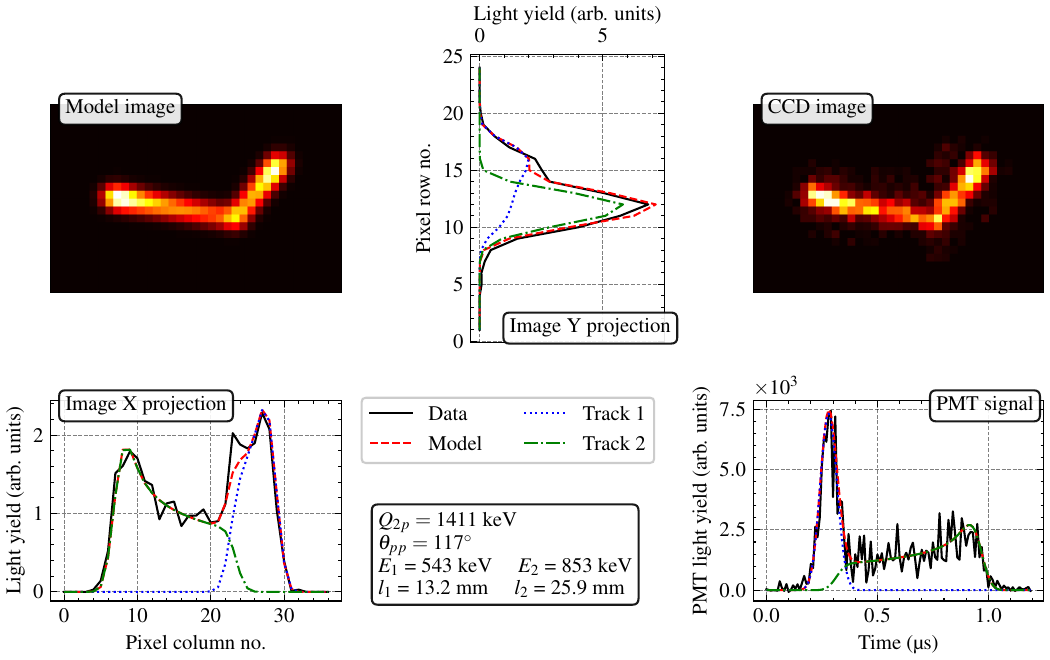}
  \caption{An example of the full reconstruction of a \emph{2p} decay event of $^{54}$Zn.
    Kinetic energies of the two protons, $E_{1,2}$, and the lengths of their tracks, $l_{1,2}$
    are given together with the angle between the tracks, $\theta_{pp}$, and the total \emph{2p}
    decay energy $Q_{2p}$.}
  \label{fig:2pReconstruction}
\end{figure*}

In Fig.~\ref{fig:bpReconstruction} an example of a $\beta$-delayed proton event is shown together
with the best-fitted model. The given kinetic energy of the proton ($E_0$) and its emission
angle with respect to the horizontal plane ($\theta$) were calculated from the best-fit parameters.
In Fig.~\ref{fig:2pReconstruction} an example of a fully reconstructed \emph{2p} event is presented. In some events, like the one shown here, more than one possibility exist which part of the waveform corresponds to which track on the image. In such cases all the combinations were considered and the best fitting one was chosen. The final results, which are the energies of the two protons and the angle
between their tracks, as measured in the plane defined by them, were calculated from best-fit parameters.

\section{Results}
\subsection{$^{56}$Zn}
The first setting of the BigRIPS separator was tuned for the optimal transmission of
$^{56}$Zn. The main purpose of this setting was to verify the tuning, collect data for
the cross-section determination, and to calibrate the OTPC detection system.
However, in addition, we could observe decays of $^{56}$Zn with delayed emission
of protons. In total 997 ions of $^{56}$Zn were identified arriving to the OTPC
detector. Out of this number, 521 were properly stopped within the active volume
of the chamber. Inspection of latter events revealed 375 $\beta p$ decays.
The maximum likelihood method yielded the half-life $T_{1/2} = (28.2 \pm 2.2)$~ms.
This value agrees within the $3 \sigma$ limit with the results obtained previously 
by Dossat et. al ($30.0 \pm 1.7)$~ms \cite{Dossat:2007} and by Orrigo et al. ($32.9 \pm 0.8)$~ms \cite{Orrigo:2016}. For the determination of the branching ratio
we took the half-life value from Ref.~\cite{Orrigo:2016} which has the smallest
uncertainty. The result for the $\beta p$ decay channel of $^{56}$Zn is:
$BR(^{56}{\rm Zn};\beta p) = (78.9 \pm 2.2$)\% This value is consistent within 3$\sigma$ with the results
published in Ref.~\cite{Dossat:2007} and Ref.~\cite{Orrigo:2016} which are
$(86.0 \pm 4.9)$\% and $(88.5 \pm 2.6)$\%, respectively.

\subsection{$^{55}$Cu}
In the setting on $^{56}$Zn also $^{55}$Cu ions were coming to the final focus.
They were much more abundant and some of them fulfilled the triggering condition.
As a result 3270 ions of $^{55}$Cu were identified as triggering the OTPC acquisition.
From this number, 1390 ions were found to be stopped inside the active detector volume.
Inspecting these events, we found 46 decays with emission of delayed proton.
Analysis of the decay time values
yielded the half-life $T_{1/2} = 44^{+18}_{-10}$~ms. This can be compared with the
values reported by Dossat et al. ($27 \pm 8$)~ms \cite{Dossat:2007},
by Tripathi et al. ($57 \pm 3$)~ms \cite{Tripathi:2013},
and by Giovinazzo et al. ($55.5 \pm 1.8$)~ms \cite{Giovinazzo:2020}.
Our result has large error bars, stemming from the small statistics, but is
consistent with the literature values, in particular with the most recent ones.
In the branching ratio calculation we adopted the most precise value from
Ref.~\cite{Giovinazzo:2020}. For the probability of the $\beta p$ decay channel
of $^{55}$Cu we obtained the value $BR(^{55}{\rm Cu};\beta p) = (4.3 \pm 0.6)$\%
This result is significantly smaller than $(15.0 \pm 4.3)$\% which is the only
value published previously \cite{Dossat:2007}.
We note that in Refs. \cite{Tripathi:2013} and \cite{Giovinazzo:2020}
no observation of delayed protons following decay of $^{55}$Cu was reported.

\subsection{$^{55}$Zn}
\begin{figure}[!ht]
  \includegraphics{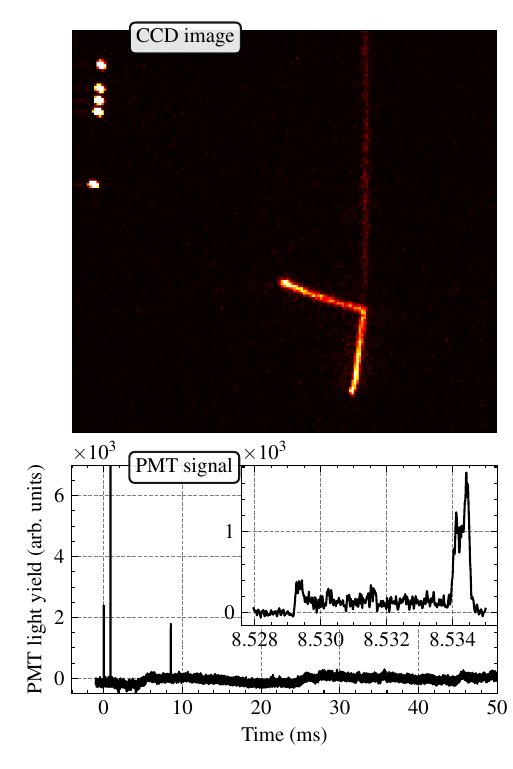}
  \caption{An example event of a $\beta$-delayed two proton emission of $^{55}$Zn.
   The CCD image is shown in the top. In the upper left part, light from the row
   of light diodes is seen which indicates the event number in the binary code.
   The bottom part shows the part of the corresponding PMT waveform. A light pulse
   at zero time originates from the triggering ion. A larger pulse next to it
   contains a series of light diode flashes which are seen merged at this scale.
   The closer inspection reveals the same bit pattern as in the CCD image.
   The simultaneous emission of two delayed protons occurred at about 8.534 ms
   after the trigger. The insert shows the zoomed part of the waveform around
   the decay time, exposing contibutions from the two protons.
  }
  \label{fig:beta2pExample}
\end{figure}

The second major setting of the BigRIPS separator was tuned for the optimal transmission
of $^{55}$Zn. Its purpose was to secure an intermediate step towards the final,
most exotic zinc isotope, and to collect data for the cross-section measurements.
Independently, we could obtain information on $\beta$ decay of $^{55}$Zn with
emission of delayed protons.
In total, 747 identified ions of $^{55}$Zn triggered the OTPC acquisition system.
303 of them were stopped within the active volume. In 269 events, decay
signals were observed. From this number, 250 events represented $\beta$ decay with
emission of a single proton ($\beta p$) and 19 events displayed delayed emission
of two protons at the same time, which are interpreted as $\beta 2p$ decays.
One such event is shown in Fig.~\ref{fig:beta2pExample}.
We note that this decay mode is observed for the first time in $^{55}$Zn.
First, we verified that the two decay modes, when analysed separately,
lead to the same half-life value within the error bars. Then, to provide a
better accuracy, we used all 269 decay events for the determination of the half-life.
The maximum likelihood analysis yielded the value $T_{1/2} = 17.9^{+1.3}_{-1.1}$~ms.
The only previous measurement available in literature, based on somewhat larger
statistics (500 events), reported the value $(19.8 \pm 1.3)$~ms \cite{Dossat:2007}.
For the branching ratio calculation, we took the half-life value determined by us.
In this case, however, the correction for the finite observation time is so small
that the half-life value from Ref.~\cite{Dossat:2007} leads to essentially the same results.
For the $\beta p$ decay channel, we get the branching ratio
$BR(^{55}{\rm Zn};\beta p) = (84.6 \pm 2.3)$\%. For the $\beta 2p$ decay channel the
resulting branching ratio is $BR(^{55}{\rm Zn};\beta 2p) = (6.4 \pm 1.4)$\%.
In the work of Dossat et al. \cite{Dossat:2007}, the branching ratio for
$\beta$-delayed protons emission was determined to be ($91.0 \pm 5.1$)\%.
This measurement was based on the energy spectrum of charged particles
recorded by a set of silicon detectors, where delayed single proton emission
could not be distinguished from delayed two-proton emission. We note that the
sum of the branching ratios for the two decay channels observed by us is equal to
the branching ratio given by Dossat et al. \cite{Dossat:2007}.

\begin{figure}[!ht]
 \includegraphics[width=1.0\columnwidth]{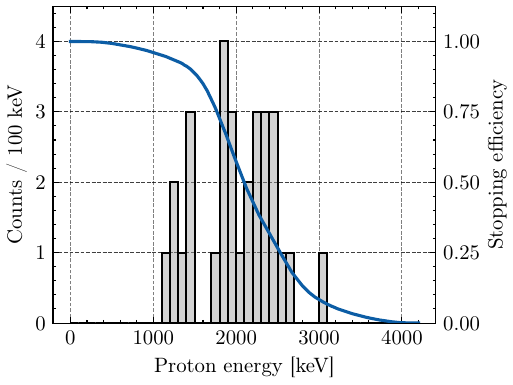}
  \caption{Energy spectrum of single protons following $\beta$ decay of $^{55}$Zn
  which were stopped in the active volume of the OTPC. The line shows the proton-stopping
  efficiency of the detector.
  }
  \label{fig:betapSpectrum}
\end{figure}

Most of the delayed protons emitted by $^{55}$Zn escaped the OTPC active volume,
so their tracks could not be reconstructed. We remind that the selected gas
mixture in the chamber was optimized for observation of low energy \emph{2p} decays.
In case of 30 $\beta p$ events, however, proton tracks were fully
contained in the detector and they could be reconstructed, as described in
section III.B, and shown in Fig.~\ref{fig:bpReconstruction}. The resulting energy
spectrum is shown in Fig.~\ref{fig:betapSpectrum}. Superimposed in this figure is
the detector efficiency as a function of energy, defined as the probability that
the proton track will be confined in the active volume. It was determined by Monte
Carlo simulations taking into account the dimensions of the detector, the
distribution of stopped ions of $^{55}$Zn, and assuming isotropic emission of
delayed protons. The spectrum of $\beta$-delayed protons from $^{55}$Zn was
previously reported only by Dossat et al.~\cite{Dossat:2007}, who measured it
with a stack of silicon detectors for energies up to 6~MeV. Unfortunately, the
stopping efficiency of the OTPC covers only the low-energy part of this
spectrum. In particular, we could not see the dominating peak at 4.7~MeV.
At low energy, we do not see any events below 1~MeV and we do not see a peak-like
structure at 1.5~MeV, visible in Fig.~55 of Ref.~\cite{Dossat:2007}. But the low
number of events in our spectrum is not sufficient for clear-cut conclusions.

From the 19 events of $\beta 2p$ emission, only one event could be fully
reconstructed and it is the event shown in Fig.~\ref{fig:beta2pExample}.
The track lengths of the two protons were found to be 49.8~mm and 103.0~mm,
corresponding to energy of 1303(51)~keV and 2056(77)~keV, respectively.

\subsection{$^{54}$Zn}
The final setting, representing the main goal of the experiment, was optimized
for the transmission of $^{54}$Zn, which is a ground state \emph{2p} emitter.
The range distribution of these ions arriving to the detection setup was found
larger than anticipated, and we recorded only five events of \emph{2p} emission.
Their decay time values yielded the half-life $T_{1/2} = 1.08^{+0.68}_{-0.37}$~ms,
which agrees within the error bars with the value $1.59^{+0.60}_{-0.35}$~ms published in Ref.~\cite{Ascher:2011}.
The five events were reconstructed as explained in Section III.B and illustrated 
in Fig.~\ref{fig:2pReconstruction}. The results are collected in Table~\ref{tab:2pResults}.
The \emph{2p} decay energy, $Q_{2p}$, was calculated as the sum of kinetic energies of both
protons and the recoil energy of the daughter nucleus which depends on the angle between
proton tracks, $\theta_{pp}$. For one of the events, the third in Table \ref{tab:2pResults},
it was not certain if both protons were fully stopped inside the active volume. This is
reflected in the larger error bars for this event. The weighted average of the \emph{2p} decay 
energy, from our five events, is $Q_{2p} = 1363(25)$~keV. This value happens to be 
almost exactly in the middle between values reported by Blank et al. (1480(20) keV) \cite{Blank:2005} and by Ascher et al. (1280(210) keV) \cite{Ascher:2011}.

\begin{table}[!ht]
\caption{Results of the reconstruction of the five $2p$ decay events of $^{54}$Zn. $E_1$ and $E_2$ are kinetic energies of both protons, $\theta_{pp}$ is the angle between their momenta, and the $Q_{2p}$ is the $2p$ decay energy.}
	\begin{ruledtabular}
		\begin{tabular}{ccccc}
		$E_1$ [keV] & $E_2$ [keV] & $\theta_{pp}$ [deg] & $Q_{2p}$ [keV] \\
		\hline
        582(27)  & 813(35)  & 136(6)  & 1402(60) \\
        543(37)  & 853(41)  & 117(2)  & 1411(55) \\
        643(63)  & 733(83)  & 136(20) & 1376 (100) \\
        518(30)  & 818(45)  & 52(2)   & 1377 (54) \\
        408(20)  & 857(37)  & 39(2)   & 1306 (42) \\
		\end{tabular}
	\end{ruledtabular}
 \label{tab:2pResults}
\end{table}

\begin{figure}[!ht]
 \includegraphics[width=1.0\columnwidth]{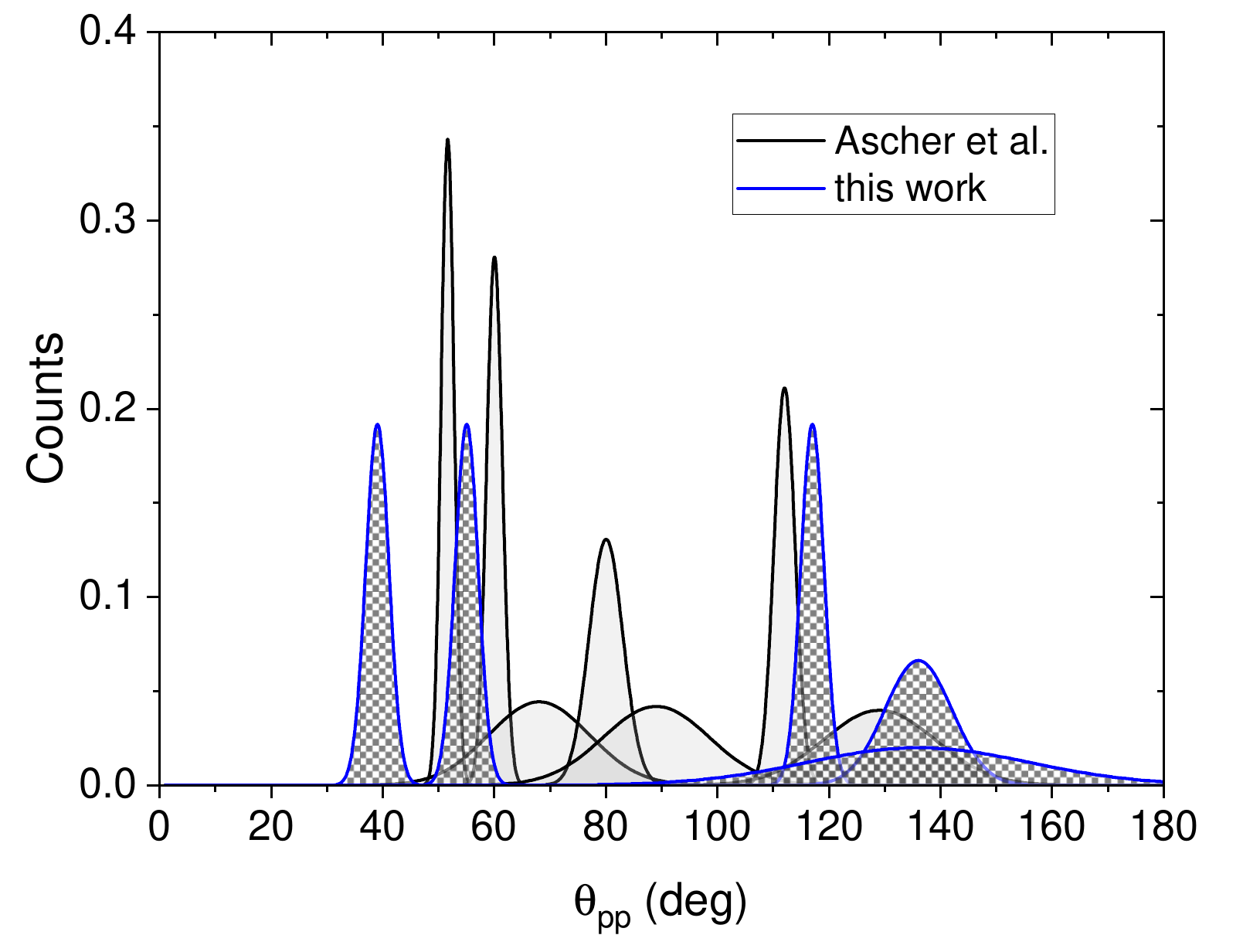}
  \caption{Opening angle between momenta of two protons emitted from the
  ground state of $^{54}$Zn. 
%  The open squares indicate values measured by Ascher et al.~\cite{Ascher:2011,Ascher:2011b} 
%  while the blue circles show the result of the present work.
   Each event is represented by the normal function with the width equal to the uncertainty
   of the angle. Black solid line indicates values measured by Ascher 
   et al.~\cite{Ascher:2011,Ascher:2011b}, while the blue lines with the pattern show
   results of the present work.
  }
  \label{fig:54Zn2pAngles}
\end{figure}

The \emph{2p} emission from $^{54}$Zn was studied before in detail by Ascher et al.~\cite{Ascher:2011,Ascher:2011b}. Using a TPC detector installed behind the LISE3
fragment separator at GANIL laboratory, they were able to fully reconstruct 
seven events of \emph{2p} decay. The interesting feature of the \emph{2p} emission 
is the correlation between the proton momenta, and in particular the opening
angle between them. The observations made for the \emph{2p} decay
of $^{45}$Fe \cite{Miernik:2007b,Miernik:2009} suggested that the angular correlation between
protons carries some information about the wavefunction of the decaying state.
Since $^{45}$Fe and $^{54}$Zn are placed on both sides of the $Z = 28$ shell
closure, they should exhibit a different proton correlation pattern, provided
the shell gap does not vanish for these extremely neutron-deficient nuclei.
The opening angle between protons emitted by $^{54}$Zn, as reported by
Ascher et al.~\cite{Ascher:2011,Ascher:2011b}, and determined by us 
(shown in Table~\ref{tab:2pResults}), are collected in Fig.~\ref{fig:54Zn2pAngles}.
Our values indicate a distribution consistent with the results of 
Ref.~\cite{Ascher:2011,Ascher:2011b}.
Although the total statistics of both experiments is not sufficient for
meaningful conclusions, we may note that the data indicate a rather flat
angular distribution. If this indeed will prove to be true, the
pattern for $^{54}$Zn would be significantly different from the one observed 
for $^{45}$Fe, in which there are four times more events with angles between 
0 and 90 degrees than with angles between 90 and 180 degrees \cite{Miernik:2007b}.
Clearly, data for $^{54}$Zn with much larger statistics are needed to verify this supposition.

\section{Summary}

Using fragmentation reactions of a $^{78}$Kr beam at 345~MeV/nucleon on 
a 10~mm-thick beryllium target, and the BigRIPS separator we produced 
nuclei in vicinity of $^{54}$Zn which is the ground-state two-proton emitter 
and the most neutron-deficient isotope of zinc known to date.
The production cross sections measured in this experiment were already
published elsewhere \cite{Kubiela:2021}. Here, we presented results of
decay studies of these nuclides. We used the Warsaw gaseous time projection 
chamber with optical readout (OTPC), installed at the final focus of the
Zero Degree Spectrometer, to record decays with emission of protons 
for nuclei implanted into the active part of the chamber. 
For $^{56}$Zn we detected 375 $\beta p$ decays which allowed us to 
determine the half-life $T_{1/2} = (28.2 \pm 2.2)$~ms and the branching
ratio $BR(^{56}{\rm Zn};\beta p) = (78.9 \pm 2.2$)\%. Both these values
are consistent with the results in the literature \cite{Dossat:2007,Orrigo:2016}.
For $^{55}$Cu we observed 56 decays with emission of a delayed proton.
The deduced half-life of $T_{1/2} = 44^{+18}_{-10}$~ms agrees 
with the most recent literature value \cite{Giovinazzo:2020}. 
The branching ratio, however, $BR(^{55}{\rm Cu};\beta p) = (4.3 \pm 0.6)$\% 
was found to be three times smaller than the value published in Ref.~\cite{Dossat:2007}.
In case of $^{55}$Zn we recorded 250 events representing $\beta p$ decay channel.
In addition, for the first time, we observed 19 events of $\beta$-delayed
two-proton emission. The branching ratios for these two channels were
determined to be $BR(^{55}{\rm Zn};\beta p) = (84.6 \pm 2.3)$\% and
$BR(^{55}{\rm Zn};\beta 2p) = (6.4 \pm 1.4)$\%. The sum of these values
coincides with the branching ratio published previously by 
Dossat et al. \cite{Dossat:2007} who could not distinguish the delayed single-proton
emission from the delayed two-proton emission. For the half-life of $^{55}$Zn
we found the value $T_{1/2} = 17.9^{+1.3}_{-1.1}$~ms which is slightly smaller
than the result of Ref.~\cite{Dossat:2007} but agrees with it within 2$\sigma$.
Finally, we could observe five events of \emph{2p} radioactivity of $^{54}$Zn. 
The distribution of the opening angles between the two protons, together with
the data from seven events measured by Ascher et al.~\cite{Ascher:2011}, 
suggests a flat distribution. If confirmed, this would differ significantly
from the pattern observed in the \emph{2p} decay of $^{45}$Fe \cite{Miernik:2007b}.
Measurements of \emph{2p} radioactivity of $^{54}$Zn, and in addition of $^{48}$Ni,
with higher statistics are needed to answer the question whether the $Z=28$
shell closure does have an impact on the \emph{2p} correlation pattern.

\begin{acknowledgments}

  We would like to thank the whole RIBF accelerator staff
  for the support during the experiment and for maintaining
  excellent beam conditions.

  This work was partially supported by the National Science Center, Poland, under
  Contracts No. UMO-2015/17/B/ST2/00581, 2019/33/B/ST2/02908, and 2019/35/D/ST2/02081,
  by the University of Warsaw Integrated Development Programme (ZIP),
  co-financed by European Social Fund within Knowledge Education Development Programme 2014-2020, p.3.5,
 % by the U.S. National Science Foundation under Grant No. PHY-2012040,
  by the Office of Nuclear Physics, U.S. Department of Energy
  under Award No. DE-FG02-96ER40983 (UTK), by the National
  Nuclear Security Administration under the Stewardship Science
  Academic Alliances program through DOE Award No. DE-NA0003899,
  and by the MEXT/JSPS KAKENHI grants No. 16K05390, 18H03602, and 18H05462.

\end{acknowledgments}

%\bibliography{MPF_pDripLine}
%\bibliography{d:/Pfutzner/Projects/BibTeX/MPF_pDripLine}
\providecommand{\noopsort}[1]{}\providecommand{\singleletter}[1]{#1}%

\end{document}